\documentclass[12pt]{article}
\usepackage{epsfig}
\newcommand{\be}{\begin{equation}}
\newcommand{\ee}{\end{equation}}
\newcommand{\bra}{\langle}
\newcommand{\ket}{\rangle}
\newcommand{\bea}{\begin{eqnarray}}
\newcommand{\eea}{\end{eqnarray}}

\begin{document}
\title{Simulations of financial markets in a Potts-like model}

\author{Tetsuya Takaishi${}$ \\
\small \it CERN, Physics Department, TH Unit, CH-1211 Gen\`eve 23, Switzerland \\
\small \it Hiroshima University of Economics, Hiroshima 731-0124, Japan \\
}

\maketitle

\abstract{ 
A three-state model based on the Potts model is proposed to simulate financial markets.
The three states are assigned to "buy", "sell" and "inactive" states.
The model shows the main stylized facts observed in the financial market: fat-tailed distributions
of returns and long time correlations in the absolute returns.
At low inactivity rate, the model effectively reduces to the two-state model of Bornholdt 
and shows similar results to the Bornholdt model.
As the inactivity increases, we observe the exponential distributions of returns. 
}

\vspace{5mm}

\section{Introduction}
Recent empirical studies of financial prices such as stock indexes, exchange rates
established the stylized facts for price returns\cite{Cont}:
fat tailed distributions, absence of autocorrelations in return, long time correlation in absolute return, etc.
A variety of models have been proposed to imitate the real financial market 
and to understand the origins of the stylized facts\cite{Zhang}-\cite{Bornholdt}.
Many models are able to capture some of the stylized facts.
Among them, simple models may serve to find the universality behind the financial market\cite{Iori}.

Bornholdt proposed a particularly simple spin model for financial markets based on the Ising model\cite{Bornholdt}.
His model has been further investigated by several authors\cite{Yamano,Yamano2,Kaizoji,Theodosopoulos}.
In the Bornholdt model the two states of the Ising spins are assigned to "buy" or "sell" orders. 
The essential ingredients of the model are two conflicting interactions: 
(i) nearest neighbor interaction (ferromagnetic) (ii) global interaction (anti-ferromagnetic).
The nearest neighbor interaction gives an effect to imitate neighbors. 
On the other hand, the global interaction coupled to the magnetization  
plays a role of the minority effect. These interactions cause a complicated dynamics, i.e., 
there is no single stable phase, rather the model exhibits metastable phases.
As a result, the distribution of the price returns shows a fat-tailed distribution
and the volatility or the absolute return 
is strongly clustered which results in the long time correlation.

Motivated by the two-state Bornholdt model, we propose a three-state model based on the Potts model.
Compared to the Bornholdt model, the model includes an additional state 
which may be assigned to "inactive" state. 
Therefore the model incorporates three states assigned to "buy", "sell" and "inactive". 
We simulate the model with the additional inactive state and  
study whether the results show the stylized facts observed in the real financial market.

\section{Potts-like Model}
Following the heat bath dynamics proposed by Bornholdt\cite{Bornholdt},
we construct a model based on the three-state Potts model.
Each agent $i$ lives on a site of $d$-dimensional lattice
and has a spin $S_i$.
First let us consider the Bornholdt model.
Spins $S_i$ have two states $(+1,-1)$ and are updated to $S_i^\prime$ with the following probabilities
\footnote{Here we use the two-state Potts representation rather than the Ising one.}.
\be
P(S_i \rightarrow S_i^\prime) = \exp(\beta(h_{i,S_i^\prime}-\alpha S_i S_i^\prime|M|))/Z
\ee
where 
\be
h_{i,S_i^\prime}=\sum_{<i,j>} \delta_{S_j,S_i^\prime},
\label{eq:h}
\ee
and the summation $\bra i,j\ket$  
is taken over the nearest neighbors $j$ of the site $i$.
$Z$ is the normalization factor determined such that the following equation is satisfied. 
\be
\sum_{k=1,-1}P(S_i \rightarrow k)=1.
\ee
$M$ is the magnetization given by
\be
M=\frac1N \sum^N_i S_i,
\ee
where 
$N$ is the total number of agents.

The factors $\beta h_{i,S_i^\prime}$ represent
the nearest neighbor interactions according to which agents imitate their neighbor.
On the other hand, the factors $\alpha S_i  S_i^\prime|M|$ 
give the minority effects. Namely if the magnetization $|M|$ gets large,
the agents tend to change their states.

Next we add an additional state corresponding to the inactive one.
We assign "0" to this state.
We define the probability to have this inactive state as 
\be
P(S_i \rightarrow S_i^\prime=0) = \exp(\beta(h_{i,S_i^\prime}-\epsilon_i\alpha |M|-2\gamma K))/\bar{Z},
\ee
where $K$ is the inactivity rate given by
\be
K=\frac1N \sum^N_i \delta_{S_i,0},
\ee
and $\epsilon_i = 2\delta_{S_i,0}-1$ which takes 1 for $S_i=0$ and $-1$ for $S_i\neq 0$.
The factor $\epsilon_i\alpha |M|$ corresponds to the minority effects 
and $2\gamma K$ controls the inactivity rate.
By varying $\gamma$ we study the model for various inactivity rates.

In summary, our updating scheme for the three-state model is as follows.
We update spins $S_i$ to $S_i^\prime$ with the following probabilities given as
\be
P(S_i \rightarrow S_i^\prime) = 
\left\{ \begin{array}{ll} 
 \exp(\beta(h_{i,S_i^\prime}-\alpha S_i S_i^\prime|M|))/\bar{Z} &{\rm for } \hspace{3mm} S_i^\prime=1,-1\\
 \exp(\beta(h_{i,S_i^\prime}-\epsilon_i\alpha |M|-2\gamma K))/\bar{Z}  &{\rm for } \hspace{3mm} S_i^\prime=0\\
\end{array} \right.
\label{eq:Pob}
\ee
where $\bar{Z}$ is the normalization factor determined by
\be
\sum_{k=1,-1,0}P(S_i \rightarrow k)=1.
\ee

\section{Results}
In our simulations, we use a $100\times 100$ square lattice with the periodic boundary condition.
When we update spins, we choose a site randomly and update it with the probabilities given by eq.(\ref{eq:Pob}).
After $100^2$ chosen sites are updated, 
time is incremented and we also call this one sweep.
The simulations are started with a random configuration.
The first $10^4$ sweeps are discarded as
thermalization and we accumulate $10^6$ sweeps for analysis. 
We use the random number generator in Ref.\cite{Makino} for a single processor of SGI Altix3700.

The simulations were done at $\beta=4$ and $\alpha=10$.
This value of $\beta$ is larger than the critical coupling $\beta_c \sim 1.005$\cite{Wu} 
of the three-state Potts model on a 2-dimensional lattice.
In other words, the temperature $T=1/\beta$ is below the critical temperature.
We define the return through the magnetization\cite{Kaizoji} as 
$r(t)=M(t+1)-M(t)$ where $\displaystyle M(t)=\frac1N \sum^N_i S_i(t)$.

In our model $\gamma$  controls the activity rate of agents.
Fig.1 shows the average inactivity rate $\bra K \ket$ as a function of $\gamma$.
As $\gamma$ increases, the inactivity decreases.
For $\gamma \ge 1$ the inactivity rate is very small: for $\gamma \ge 1$ 
the inactivity rate $\le 10^{-3}$.
Thus for $\gamma \ge 1$ the model is expected to effectively reduce to the two-state model similar to the Bornholdt model.

Fig.2 shows the distributions of returns at $\beta =4$ and $\alpha=10$ for various $\gamma$.
The distributions are not the Gaussian distributions but shows fat tailed ones.
The shape of the distribution at $\gamma=1$  
is similar to that observed previously in the Bornholdt model\cite{Kaizoji}. 
As $\gamma$ decreases, however, the distribution changes its shape.
At $\gamma \approx 0.05$ the shape seems to be an exponential.
We will study the shape of the distribution in detail with the cumulative distributions. 

Fig.3 shows the cumulative distributions of the absolute returns $|r|$.
For $\gamma=1.0$, the power-law behavior $\sim r^{-\nu}$ is seen in a range of $0.002<r<0.1$.
The exponent $\nu$ is estimated to be $\alpha = 2.3$ which is 
the same value obtained in Ref.\cite{Kaizoji}.
This is not surprising because at this $\gamma$ the model is effectively 
a two-state model.  

As $\gamma$ decreases, the power-law behavior disappears and 
the shape of the distribution seems to change to the exponential at large $r$.
Assuming the exponential form $\exp(- \eta r/\sigma_r)$, 
we obtain $\eta=1.07$ at  $\gamma=0.05$ for $r>0.004$.
$\sigma_r$ is the standard deviation of $r$, $\sigma_r^2 =\bra (r-\bra r\ket)^2\ket$.
Recent studies claim that the distributions of returns in emerging markets show
the exponential distribution. The values of $\eta$ are evaluated 
to be $\eta \approx 1.3$ and 1.5 in the Indian market\cite{Stanley}, and
1.7 in the Brazilian market\cite{Brazil}. 
Furthermore there is also a claim that 
at mesoscopic time scale the distributions in developed markets  
also show the exponential distributions\cite{Yakovenko}.

The autocorrelation function $C(t)$ of a time series $g(k)$ is defined as
\be
C(t)=\frac{\bra g(k)g(k+t)\ket -\bra g(k)\ket^2 }{\sigma_g^2},
\ee 
where $\sigma_g^2 = \bra g(k)^2\ket -\bra g(k)\ket^2$.
Fig.4 shows the autocorrelation functions $C(t)$ of returns.
As observed in the real financial market, the autocorrelation of returns
disappears quickly as the time increases.
An interesting observation in Fig.4 is that the autocorrelation is negative 
at small $t$.
This negative autocorrelation is also seen in the financial market\cite{Cont}.

Contrary to the autocorrelation of the returns, the autocorrelation of the absolute returns
shows long time correlation.
Fig.5 shows the autocorrelation of the absolute returns for $\gamma=1.0$ and 0.05.
At $\gamma=1.0$, we do not see a power-law behavior but rather observe an exponential behavior.
Assuming $\exp(-t/t_0)$, we obtain $t_0 \approx 1280$ or $1/t_0\approx7.8\times 10^{-4}$  
by a fit in a range of $ 400 \le t \le 3000$.
Thus the autocorrelation function is a very slowly decaying exponential.
The similar exponential behavior of the autocorrelation function is also seen in the Bornholdt model\cite{Yamano}.  
As $\gamma$ decreases, the autocorrelation starts to show a power-law behavior.
At $\gamma=0.05$ we observe $C(t)\sim t^{-0.34}$. 

\begin{figure}
\vspace{5mm}
\begin{center}
\epsfig{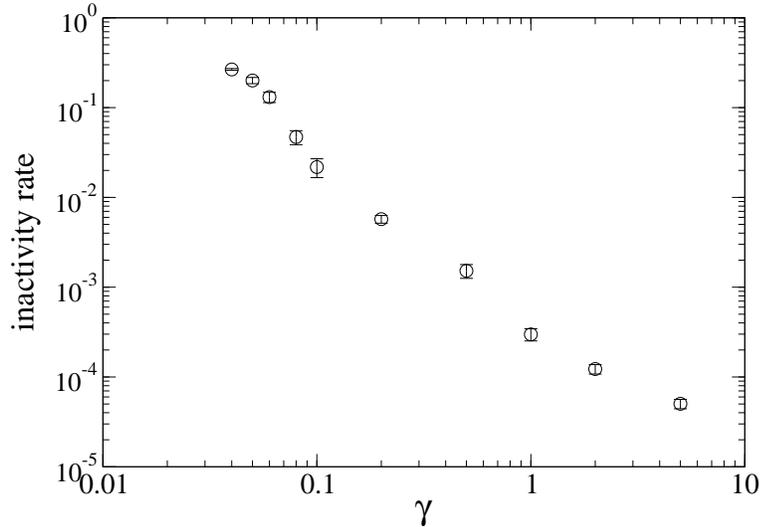}
\caption{\label{fig:inactive} 
Average inactivity rate at $\beta=4$ and $\alpha=10$ as a function of $\gamma$.
}
\end{center}
\end{figure}

\begin{figure}
\vspace{5mm}
\begin{center}
\epsfig{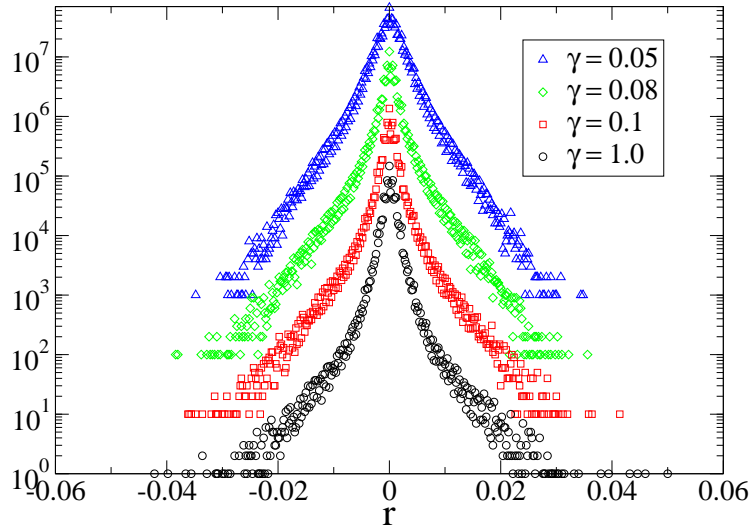}
\caption{\label{fig:hist}Histogram of returns at $\beta=4$ and $\alpha=10$ for various $\gamma$. 
Each histogram except at $\gamma=1.0$ is shifted up adequately to avoid overlapping each other.
}
\end{center}
\end{figure}

\begin{figure}
\vspace{5mm}
\begin{center}
\epsfig{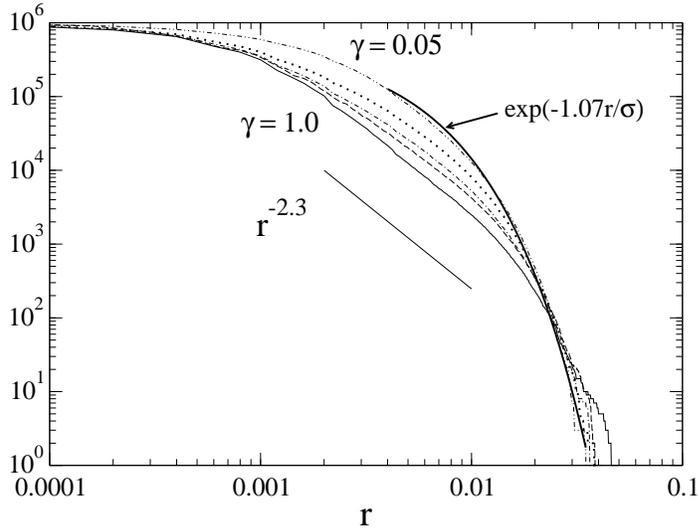}
\caption{\label{fig:cum} Cumulative histograms for $\gamma=1.0,0.5,0.1,0.08,0.05$ from bottom to up.
A line proportional to $r^{-2.3}$ and a curve of $\exp(-1.07 r/\sigma_r)$ are also drawn.
}
\end{center}
\end{figure}

\begin{figure}
\vspace{5mm}
\begin{center}
\epsfig{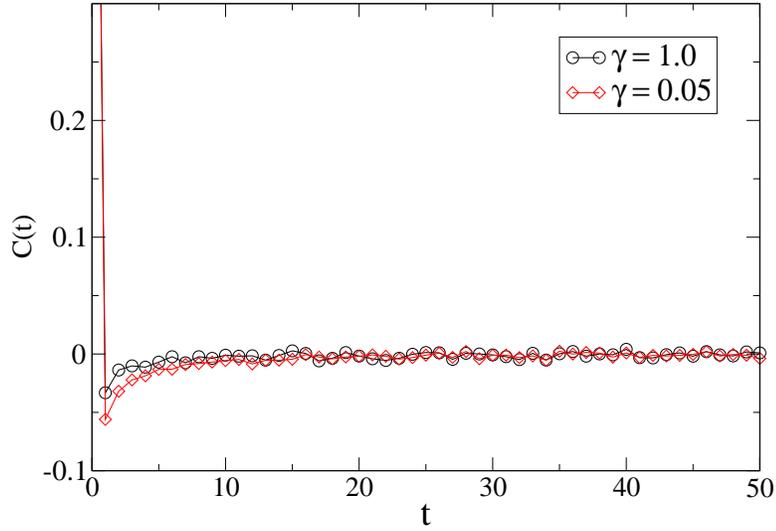}
\caption{\label{fig:auto} 
Autocorrelation function of the returns for $\gamma=1.0$ and 0.05.
}
\end{center}
\end{figure}

\begin{figure}
\vspace{5mm}
\begin{center}
\epsfig{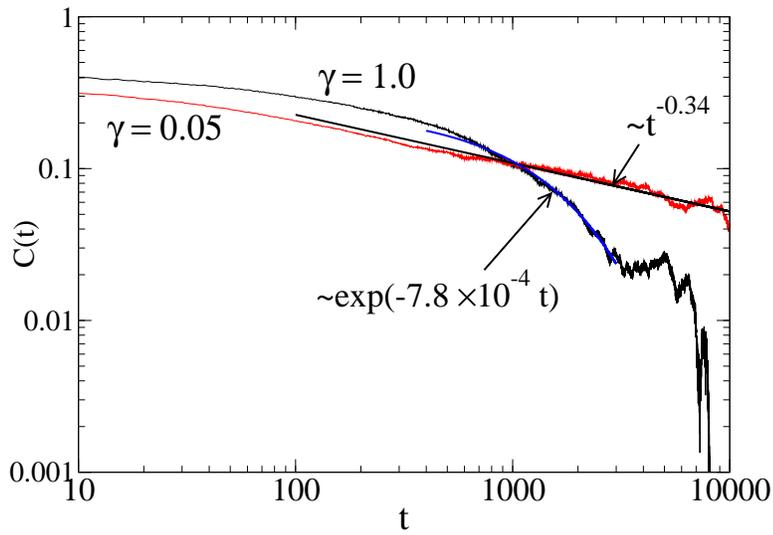}
\caption{\label{fig:abauto} 
Autocorrelation function of the absolute returns for $\gamma=1.0$ and 0.05.
}
\end{center}
\end{figure}

\section{Summary}
We proposed a three-state model based on the Potts model.
The results shows the main stylized features: fat-tailed distributions
of returns, long time correlations in the absolute returns.

At low inactivity rate $(\gamma \ge 1)$ the model 
effectively reduces to the two-state model 
similar to the Bornholdt model.
We observed a power-law behavior for the cumulative distribution of the returns:
$\sim r^{-2.3}$.
The autocorrelation of the absolute returns does not show a power-law behavior
but an exponential behavior.

As the inactivity increases, the power-law behavior for the cumulative distribution of the returns 
disappears and 
the cumulative distribution shows an exponential behavior.
Although the empirical study of the financial markets shows a power-law\cite{Cubic} $\sim r^{-3}$,
the exponential behavior observed in our model is not peculiar.
Recent studies\cite{Stanley,Brazil} claim that emerging markets such as those in Brazil and India 
show the exponential distributions of returns.
Moreover, the distributions of the developed markets at mesoscopic time scale also 
show exponential behavior\cite{Yakovenko}.

\section*{Acknowledgements}
The author thanks the Institute of Statistical Mathematics for the use of SGI Altix3700. 
He is grateful to Ph. de Forcrand for valuable comments.
He would like to thank Prof.Sigrist for hospitality during his stay at ETH Z\"urich.


\begin{thebibliography}{99}
\bibitem{Cont}
R.~Cont, Quantitative Finance {\bf 1}, 223 (2001)
\bibitem{Mantegna}
R.~Mantegna and H.E.~Stanley, Introduction to Econophysics (Cambride University Press) (1999)
\bibitem{Zhang}
D.~Challet, A.~Chessa, M.~Marsili and Y-C.~Zhang, Quantitative Finance {\bf 1}, 168 (2001)
\bibitem{Raberto}
M.~Raberto, S.~Cincotti, S.M.~Focardi and M.~Marchesi, Physics A {\bf 299}, 319 (2001)
\bibitem{CB}
R.~Cont and J.-P.~Bouchaud, Macroecon. Dynamics {\bf 4}, 170 (2000)
\bibitem{Penna}
D.~Stauffer and T.J.P.~Penna, Physics A {\bf 256}, 284 (1998)
\bibitem{Silva}
L.R.~da Silva and D.~Stauffer, Physics A {\bf 294}, 235 (1998)
\bibitem{Lux}
T.~Lux and M.~Marchesi, Nature {\bf 397}, 498 (1999)
\bibitem{Iori}
G.~Iori, Int. J. Mod. Phys. C {\bf 10}, 1149 (1999)
\bibitem{Sznajd}
K.~Sznajd-Weron and R.~Weron,  Int. J. Mod. Phys. C {\bf 13}, 115 (2002)
\bibitem{Sanchez}
J.R.~Sanchez, Int. J. Mod. Phys. C {\bf 13}, 639 (2002)
\bibitem{Bornholdt}
S.~Bornholdt, Int. J. Mod. Phys. C {\bf 12}, 667 (2001)
\bibitem{Yamano}
T.~Yamano, Int. J. Mod. Phys. C {\bf 13}, 89 (2002)
\bibitem{Yamano2}
T.~Yamano, Int. J. Mod. Phys. C {\bf 13}, 645 (2002)
\bibitem{Kaizoji}
T.~Kaizoji, S.~Bornholdt and Y.~Fujiwara, Physica A {\bf 316}, 441 (2002)
\bibitem{Theodosopoulos}
M.~Badshah, R.~Boyer and T.~Theodosopoulos, math.PR/0501244; math.PR/0501249
\bibitem{Makino}
J.~Makino, T.~Takaishi and O.~Miyamura,
Comput. Phys. Commun. {\bf 70}, 495 (1992)


\bibitem{Wu}
F.Y.~Wu, Rev. Mod. Phys. {\bf 54}, 235 (1982)

\bibitem{Stanley}
K.~Matia, M.~Pal, H.~Salunkay and H.E.~Stanley, Europhys. Lett. {\bf 66}, 909 (2004)
\bibitem{Brazil}
L.~Cout Miranda and R.~Riera, Physica A {\bf 297}, 509 (2001)
\bibitem{Yakovenko}
A.~Christian Silva, R.E.~Prange and V.M.~Yakovenko, Physica A {\bf 344}, 227 (2004)
\bibitem{Cubic}
P.~Gopikrishnan, M.~Meyer, L.A.N.~Amaral and H.E.~Stanley
Eur. Phys. J. B {\bf 3}, 139 (1998)


\end{thebibliography}
\end{document}